# A Crowdsourced Contact Tracing Model to Detect COVID-19 Patients using Smartphones

Linta Islam [a,*], Mafizur Rahman [b], Nabila Ahmad [b], Tasnia Sharmin [b], Jannatul Ferdous Sorna [b]

[a] *Department of CSE, Jagannath University, Dhaka, Bangladesh*
[b] *Department of CSE, East West University, Dhaka, Bangladesh*
Corresponding author: *[*]linta@cse.jnu.ac.bd*

*Abstract*— **Millions of people have died all across the world because of the COVID-19 outbreak. Researchers worldwide are working together and facing many challenges to bring out the proper vaccines to prevent this infectious virus. Therefore, in this study, a system has been designed which will be adequate to stop the outbreak of COVID-19 by spreading awareness of the COVID-19 infected patient situated area. The model has been formulated for Location base COVID-19 patient identification using mobile crowdsourcing. In this system, the government will update the information about inflected COVID-19 patients. It will notify other users in the vulnerable area to stay at 6 feet or 1.8-meter distance to remain safe. We utilized the Haversine formula and circle formula to generate the unsafe area. Ten thousand valid information has been collected to support the results of this research. The algorithm is tested for 10 test cases every time, and the datasets are increased by 1000. The run time of that algorithm is growing linearly. Thus, we can say that the proposed algorithm can run in polynomial time. The algorithm's correctness is also being tested where it is found that the proposed algorithm is correct and efficient. We also implement the system, and the application is evaluated by taking feedback from users. Thus, people can use our system to keep themselves in a safe area and decrease COVID patients' rate.**

*Keywords*—**Contact tracing; crowdsourcing; Covid-19; location tracking; mobile application.**

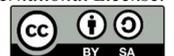

## I. INTRODUCTION

Crowdsourcing is a platform that includes getting a task, valuable information, or ideas from a broad group of people. People can easily earn money as paid freelancers and do small voluntary functions on this platform [1]. Mobile crowdsourcing allows people to gather and process a large amount of data through data sharing infrastructure over third parties. Mobile crowdsourcing can quickly bring out unexpected solutions for several challenging issues [2]. Therefore, mobile crowdsourcing for healthcare services renders a way to bring out a new way of life to lead a healthy life by identifying several diseases and their solutions.

According to Statista, the number of smartphone users increased sharply to 3.2 billion from 2016 to 2021 [3]. Hence, mobile crowdsourcing is becoming more accessible to the world through internet devices like smartphones or tablets because of the technological revolution. As these devices provide reliable GPS, therefore, through smartphones or tablets, one can trace others' location, which makes mobile crowdsourcing more convenient and efficient [4]. This location-based crowdsourcing allows workers who have mobile phones to easily trace the task requester's location [5]. Also, the crowd workers do not have to take hassles to complete any task within the accounted time as they can work on the exact location by physical presence.

At present, the world is suffering a major disaster because of a new coronavirus. This virus is transmitted into human bodies through droplets from the coughs, exhales, and sneezes produced by an infected person [6]. As the droplets are very heavy, they cannot hang on the air and immediately fall on the surfaces. One can be affected by breathing in the virus if they stay within the proximity of an infected person or contaminated surfaces [7]. The infected person often has some symptoms like dry coughs, fever, and tiredness. Worldometer reports over one hundred seven million cases, and more than two million people worldwide already died from this infectious disease [8]. Thus, researchers worldwide are trying to find out the proper vaccine for this disease to stop the death ratio. If this epidemic could not be appropriately controlled within a short time, the world will have to face several significant crises for this disease.

Bangladesh is a developing nation with an enormous population. Already, Bangladesh's economy deteriorates dramatically, and people meet thousands of challenges for their daily livelihood. Small scales companies are going to shut down their activities. Hence, people are losing their jobs gradually [9]. Thus, it is mandatory to identify infected individuals who are staying closer to the infected person.

Tracing the infected person's contact will be an effective solution to decrease the infection rate of covid. The government can quickly determine the affected person who should stay in self-isolation or quarantine through contact tracing. However, manual contact tracing is still a challenging task for various countries. Especially in a developing country like Bangladesh, it is challenging to deploy manual contract tracing using government forces because of lack of proper resources. Hence, developing an automated process like crowdsourced contact tracing apps will be a steppingstone for developing nations.

We propose a novel framework of contact tracing of COVID patients through mobile crowdsourcing in this circumstance. This paper's main motive is to locate COVID patients by tracing their contact who are walking outside without maintaining their quarantine and immediately aware nearby people who are walking nearby that COVID patient. This study includes four participants, i.e., user, health service provider, telecommunication service provider, and government. As the telecom company has an extensive network, they can identify any contact by using GPS. They will send the location data to the government, marking the place as safe or unsafe. This framework can efficiently identify the infected people, reducing the amount of death ratio in the future as the people will get the notification of their nearby patients. The utilized algorithm has been justified by its effectiveness and correctness of the algorithm. Also, the app user has a chance to provide valuable feedback regarding app performance.

Several research works have already been published focusing on covid-19 through contact tracing. The past year on March 20, an app named TraceTogether was launched by the Ministry of Health in Singapore, which works via the exchange of tokens through a Bluetooth link between nearby phones [10]. A self-reported application-based tracking system for Covid-19 was tested in paper [11] where participants gave their gender, age, postal code, and previous history of medical affairs to trace the manifestations of the Novel Coronavirus. However, participants are updated regularly via push alerts to confirm if they have developed symptoms and whether they have been screened for SARS-CoV-2.

For conveying the Covid-19 victims by ambulances from one place to another, a Real-Time-GPS tracking-based app is developed and presented in paper [12], which could help the Traffic Police ensure distance from the victims to the other public. To obtain its primary purpose, many Google Services and APIs were used in the application, including Firebase, a Real-Time database from GOOGLE. Anwar *et al.* [13] observed the testing level by counting the test case number in Bangladesh, which is not adequate. The authors also imply that it is high time to upgrade the testing quality and cover geological coverage by enforcing an effective contact tracing method. However, pandemic surveillance is now focused on map operations, drawing on the GIS (Geographic Information System) environment [14].

The WHO and other health agencies have extensively used visualization of map and geographical analysis to control epidemic outbreaks over the previous 20 years. Therefore, it presents the authorities with many realistic chances to enhance control over the pandemic outbreak. Moreover, it is achieved by GISon's capability to provide fast-tracking and route identification and can also predict the location of the contaminated individuals [15]–[17].

An IoT-based app has been designed by Hossain et al. [18] in Bangladesh to track the confirmed and recovered covid patient's data globally and locally. It has a feature to help people by navigating Google Maps to find the nearest hospitals and providing helpline numbers. The South Korean authority invented a fast, accurate, and unique system with mapping and GIS tools to quickly track the affected people [19]. It was also planned for the locations where every detected patient has been before. Islam *et al.* [20] developed a mobile application to measure a person's mental stability and provide mental health care during the covid situation. The researchers conducted a semi-structured survey before deploying their apps to reveal apps requirements.

Jung and Agulto [21] focused on data privacy with hidden keys while sharing details on the dedicated apps for tracing. A software-defined networking controller-centric worldwide platform was proposed in that paper for tracking and monitoring info for the cases of Covid-19 based on real-time IDS (Information Disclosure Services) to global CSCs (Centers for Disease-Control and Prevention) and daily users. Furthermore, their software met device scalability and reduced the duration of Query/Reply, where even the platform incorporates a vast number of CDCs and people in charge per CDC globally.

As per our literature study, we found most of the apps can only trace the COVID patients [10], [18], [20], and the authors were mainly concerned about the app's data and privacy [19] of the patients. Paper [22] outlined a systemic review of mobile apps using the contact tracing model from different electronic databases like google scholar, PubMed, Google, and others. The findings confer that from 18 May 2020 to 31 May 2020, only 15 apps developed for contract tracing for covid-19. For these reasons, this study focused on creating a crowdsource-based new contact tracing system concerning users and patients' security and privacy issues. The main objectives of this study are defined below:

- To build a contact tracing framework for detecting COVID infected patients and notify nearby users from the government end.
- To use mobile crowdsourcing to collect COVID patients' information and locate them by using a smartphone sensor.
- To evaluate the system scalability by taking feedback from the users.

Corona Tracer BD [28] is the only corona tracing application available in Bangladesh, and it also includes Bluetooth connectivity. Hence, this new application will not require a Bluetooth connection. The application is not limited to trace only COVID patients but also traces the nearby person of COVID cases for sending awareness notification by the government. Therefore, it is clear that this app will be a positive addition for developing countries like Bangladesh for enhancing the COVID-19 situation by reducing COVID infection.

The rest of the paper is organized as follows. Section II describes the proposed methodology and the overall working procedure of the system. Section III evaluates the performance analysis of the system and differentiates the system from other existing methods. Section IV overviews the whole system and concludes the article.

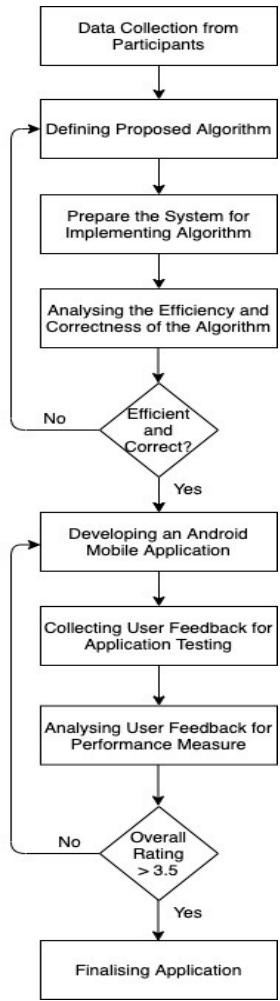

Fig. 1 Flowchart of Proposed System

## II. MATERIALS AND METHOD

This section represents a novel formal model for contact tracing for COVID-19 using mobile crowdsourcing. In this framework, the crowdsourced user could be irregular, and dynamic based on both location and time. We discuss the fundamental concepts of the proposed system and illustrate the workflow diagram briefly.

### A. Overall Working Procedure

In this subsection, we will discuss the overall working procedure of our proposed system. In Fig. 1, the flowchart of this paper has been illustrated. At the first stage, we will collect the data from different types of participants explained in sub-section II C. We will also generate some random data for our testing purpose. The second – third - fourth stage is iterative. We will finalize our proposed algorithm based on the efficiency and correctness of the algorithm. If our algorithm is efficient and correct, we will move to the next stage. In stage 5, we will develop an android application and send it to users to test our application. Based on the user feedback, we will analyze whether the application performs well or not. This fifth – sixth – seventh stage is also an iterative process. If the user feedback is below the threshold value, then we will continuously improve our application.

### B. Proposed System

Fig. 2 briefly explains our proposed system. The Government will update the information of the COVID infected patients and notify other users to stay at 6 feet or 1.8 meters distant to remain safe. We can see that the red-marked phone user is the infected user. So, the Government will mark the unsafe area making a circle where the center is the infected user. COVID might infect anyone inside the ring, so the government sends a notification to them to quickly move to the outside of that circle, denoted as the safe area.

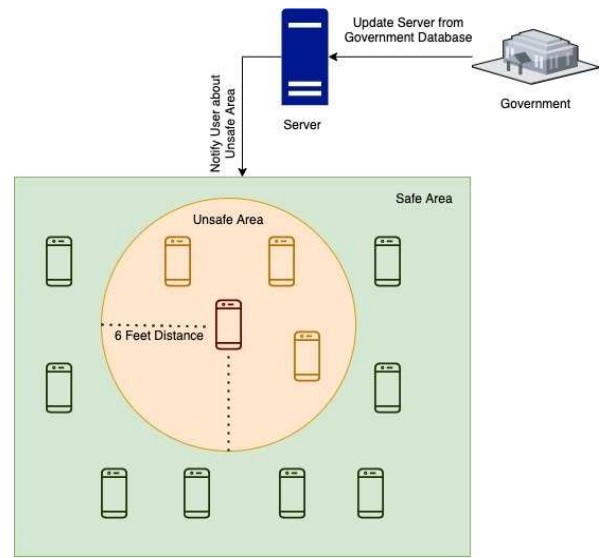

Fig. 2 System Model

### C. Types of Participants in the Model

There are four types of participants in our proposed system, which has been illustrated in Fig. 3.

*1) User:* The users are the citizen of the country and the primary consumers of our system. The set of users is denoted by *U* which includes the registered users of the system.

*2) Health Service Provider*: The health service providers are the hospitals or the diagnostic centres where a person can test COVID. The set of health service providers are denoted as *HSP* which includes the registered healthcare service providers in this system.

*3) Telecommunication Service Provider*: The telecommunication service providers are the telecom company of a country. The set of telecommunication service providers are denoted as *TSP* which includes all the registered telecommunication service providers in this system.

*4)* Government: The government is the system or a group of people who runs the country by employing proper law and legislature. The government is denoted by *G* who does all the administrative task in this system.

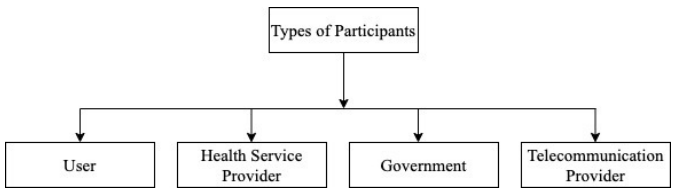
Fig. 3 Participant Types

The four types of entity will help us track a COVID patient and keep safe other people safe. This system will help reduce the infection and death rate, also increasing the recovery rate.

*D. Workflow of Proposed Model*

Fig. 4 illustrates the workflow of our proposed architecture. Firstly, the health providers will share the information of COVID patients with the government. Whenever a person tests COVID, the information will be automatically submitted to the government database by the health providers if the result is positive. The users also can offer the patient information to the government. Thus, the government database will be continuously updated. Secondly, the government will ask the telecom providers to track the location of the patient. The telecom company will follow the patient and share the location data with the government. Then the government will mark the area as unsafe using our proposed algorithm and notify all the nearby users about the hazardous area. This notification will help the user to stay away from COVID infected patients. Furthermore, the government can reduce the infection and death rate of the citizens.

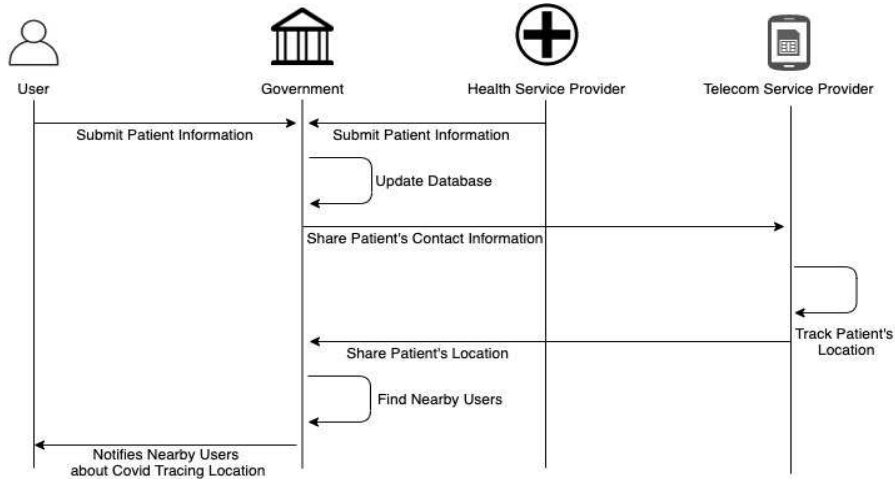
Fig. 4 Workflow of Proposed Model

*E. Unsafe Area Marking Algorithm*

The telecom service provider (TSP) will only share the location of the infected patient. Thus, we need to mark the infected area and warn the people who are near that area. TSP shares the latitude and longitude of an infected person with the government. Using algorithm 1, the government will find and mark the area as unsafe.

**Algorithm 1:** Find Unsafe Area

Input:    plat, plon
Output:   unsafe_area
1.   R = 6371
2.   PI = 3.1416
3.   dlat = plat × (PI / 180)
4.   dlon = plon × (PI / 180)
5.   a = sin(dlat/2)$^2$ + cos(plat + 1.6) × cos(plat) × sin(dlon/2)$^2$
6.   c = 2 × arctan($\sqrt{a}$, $\sqrt{(1-a)}$)
7.   d = R * c
8.   rad = (d + n)
9.   area = PI * rad$^2$
10.  **return** area

Let, plat and plon is the latitude and longitude of the infected patient that the TSP has shared. To calculate the radius of our unsafe area, we first need the earth's radius, which is 6371 km. Then, we find another safe latitude point at a distance of 1.6m from the patient. In this algorithm, the Haversine formula [20] is used to detect the patient's distance and safe point. Using this formula, we will get the distance measurement of the unsafe area. There might be some noise while calculating the distance. We need to address those noise to get the accurate distance for both safe and unsafe area. Thus, we add noise n with the length to find the radius of our unsafe area. After that, using the circle formula, we find our desired hazardous site. The government will notify everyone who is inside or near the dangerous area.

III. RESULTS AND DISCUSSION

We have implemented our proposed algorithm in the C++ language. A data set of 10000 data is created to evaluate the performance of our algorithm. The algorithm is then tested on a computer with a Core i5 2.3 GHz processor with 4GB RAM under Windows 10 64-bit operating system. We collected 10000 valid latitudes and longitudes that represent the current location of each infected patient. Using this data set, we evaluate the performance and complexity of our proposed algorithm.

*A. Computational Efficiency of the Algorithm*

At first, we measure the computational efficiency, that is, our proposed algorithm's execution time. An algorithm must provide the correct result within the polynomial time. We have tested our algorithm with 10 test cases. All the data size and runtime for each test case are illustrated in Table 1.

TABLE I
DATA SIZE VS RUN TIME OF THE ALGORITHM

| Test Case | Data Size | Runtime (in millisec) |
|---|---|---|
| 1 | 1000 | 10 |
| 2 | 2000 | 22 |
| 3 | 3000 | 30 |
| 4 | 4000 | 39 |
| 5 | 5000 | 47 |
| 6 | 6000 | 62 |
| 7 | 7000 | 71 |
| 8 | 8000 | 83 |
| 9 | 9000 | 97 |
| 10 | 10000 | 113 |

The dataset is uniformly divided into ten test cases, and each test case is merged with the previous test cases. The first test case contains 1000 data, and we increase the test data size with 1000 data in each test case. The algorithm takes ten milliseconds to find the area of 1000 patient. The runtime of the algorithm is linearly increasing with the size of the data. Finally, for 10000 patients, the algorithm takes 113 milliseconds. From the graph shown in Fig. 5, we can conclude that our proposed algorithm can run in polynomial time.

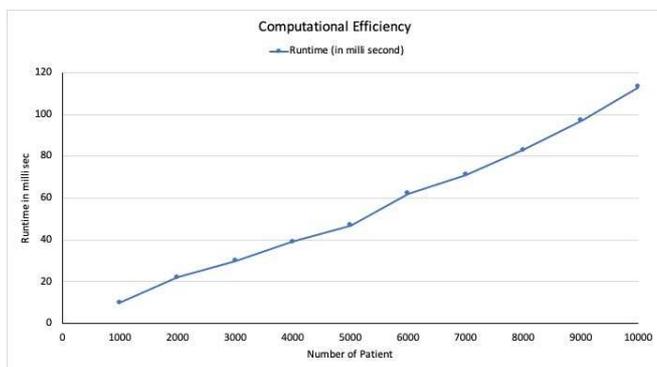

Fig. 5 Computational Efficiency of Proposed Algorithm

*B. Correctness of the Algorithm*

In the next step, we will test the correctness of the algorithm. We stored the unsafe area in another dataset for testing purpose. The latitude and longitude of a point are given as input to test whether the point is safe or unsafe. Each output is also manually checked to prove the correctness of algorithm 1. The sample output is also illustrated in Fig. 6. The test input was different types of latitude and longitude points, and the result is whether the point is in the safe area or not. This figure shows three users are in a secure location, and one user is in an unsafe place. Thus, our proposed algorithm is correct and also efficient.

```
The user in location (23.750336, 90.448566) is in safe area
The user in location (23.734135, 90.416088) is in unsafe area
The user in location (23.731984, 90.415058) is in safe area
The user in location (23.731968, 90.415090) is in safe area
```

Fig. 6 Sample Output

*C. System Implementation*

At the final stage, we have implemented the system by developing an android application. We have selected android as it is a widely used open-source framework. We used Android Studio 4.1.2 for implementing the system and Java language to write the codes. We also use Firebase to store the COVID patients' information as our database. We have adopted our smartphone's GPS (Global Positioning System) for accurate time location tracking and used several google services and APIs (Application Programming Interfaces).

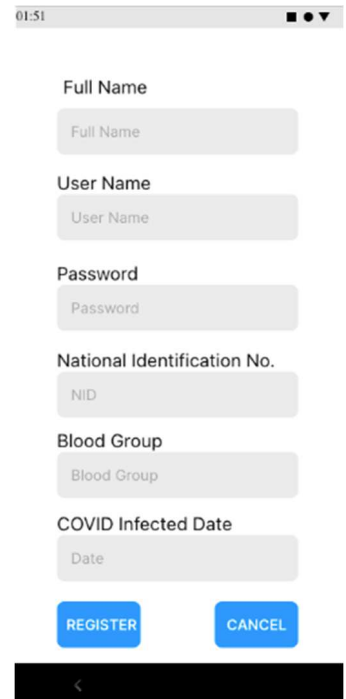

Fig. 7 Registration Interface of the System

All the users of this application should register themselves in the system. The registration interface is illustrated in Fig. 7. The users have to input their full name, username, password, National Identification Number (NID), blood group, and the date when COVID infects them. The registration process is straightforward so that the user can easily register themselves in the system.

After completion of registration, the user can search the nearby infected persons and also, they can find out whether they are in the safe area or not. In Fig. 8 (a), we can see that the user location is far from the infected person's place. Thus, he/she is in a safe area. In Fig. 8 (b), the user location is near the infected person's location. Thus, he/she is in an unsafe place. In this way, we can easily find out the location of the infected person to stay safe from COVID.

*D. System Evaluation*

We have shared our application's APK (Android Application Package) version with 110 users with a survey form to evaluate the application. However, only 62 users have submitted the survey form. There were several criteria based on which the users gave their final rating points. The requirements are (a) user-friendliness, (b)stability, (c) speed, (d) security and (e) overall rating. Fig. 9 illustrates the number of the user's feedback on each criterion.

Our first criterion is the application should be user-friendly. The user can easily register themselves in the system as well as they can comfortably operate the application. Thus, most users think the proposed approach is user-friendly by giving "Excellent" and "Good".

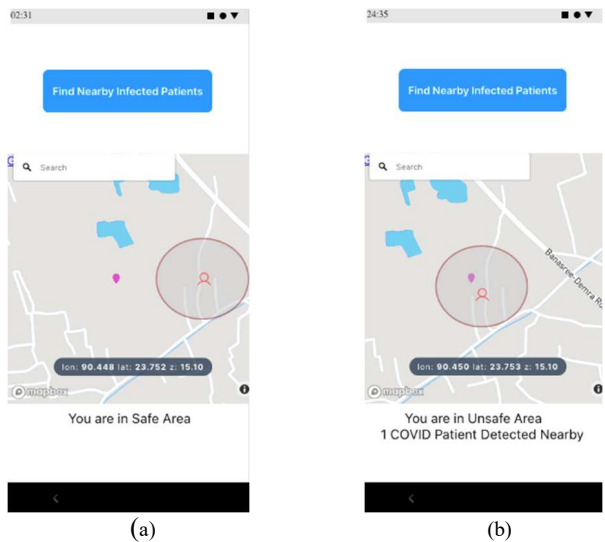

(a)            (b)

Fig. 8 Finding safe and unsafe area

Sometimes an application might freeze while running. We asked our users whether they have faced any issues related to freezing applications. Most of the users have answered that the application was stable throughout their usage time. This rating proves that our system does not freeze unexpectedly while running.

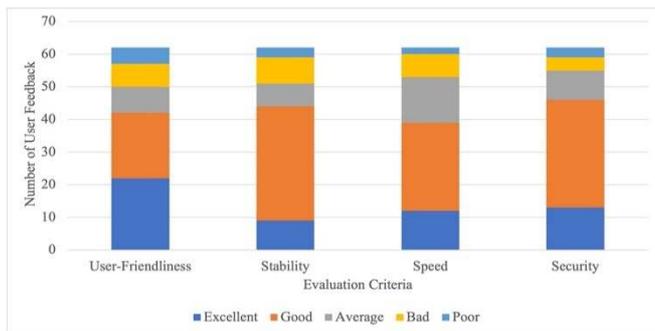

Fig. 9 Summarization of User Feedbacks

Our system is minimal in size as we do not store the device's user information and other information. Therefore, our system works faster, and it is very responsive to the user's requests. Thirty-nine users among 62 users think that our system is faster by selecting "Excellent" and "Good".

All the current works available now are based on Wi-Fi or Bluetooth technology connectivity to the best of our knowledge. Several security attacks can occur when our device is connected to another device [23-24]. For this purpose, all the existing methods are vulnerable to security attack. However, our developed application does not connect the user device to any other device. Thus, the security attack from another device cannot infect the user's devices. For this reason, 88% of users felt secure about their data while using this application.

Finally, the users provided their rating point based on their experience and satisfaction. From Fig. 10, we can see that 42 users have rated the application as "Good" and nine users as "Excellent". Only two users have rated "Poor". From this evaluation process, we can conclude that our proposed system is "Good" based on user-friendliness, stability, speed, and security.

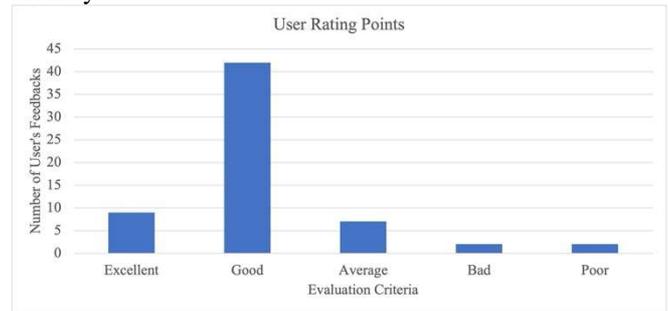

Fig. 10 Overall Rating Points of Users

### E. Discussion

The COVID-19 is spreading faster than any other virus in the world. This virus motivates us to find a solution to take preventive measure and stay away from the infection. One of the significant reasons for the virus spread was close contact with the infected patient. For this reason, we implemented an application to trace the infected people and notify others to remain safe. Several applications have already been implemented to trace contact in different countries. However, people are still unwilling to use the tracing applications for several reasons like user-friendliness, privacy, etc. Thus, we propose an easy-to-use, privacy-aware application to detect COVID patients using mobile crowdsourcing.

In this system, we use GPS to detect the patient's location and identify the unsafe area. The nearby people to this infected patient are notified through our system. Thus, this system creates awareness among people as well as prevents the expansion of COVID-19. Moreover, our system does not use Bluetooth technology like TraceTogether [10], CAUDHT [25]. One of the prime Bluetooth-based attacks is Bluejacking [26] and Bluesnarfing [27]. These attacks can cause intrusions and can access the files stored in the system. In Bangladesh, to the best of our knowledge, there is only one corona tracing application named Corona Tracer BD [28], which also needs Bluetooth connectivity to detect COVID safe distance. As our proposed system does not connect to Bluetooth to trace contacts, our system is not vulnerable to any Bluetooth-based attack. Thus, no sensitive data is shared with any other users.

Furthermore, our application uses crowdsourcing to collect information from the users, the health care provider, and the government. This data collection process increases the trust in users since there is less possibility to have misleading location information about the COVID-19 infected areas.

Despite having several benefits, our system has some limitation. Only 62 users have used our application and rated the application based on their personal experience. If we could engage more users to test this application's performance, we would have to better understand this user rating. Moreover, the user category was heterogeneous concerning gender and homogeneous concerning age and education level. All the user's feedback was collected from the student at the university level with an age-limited to 18-22 years.

This crowdsourced-based contact tracking system worked efficiently by eradicating previous limitations such as connectivity, security, complexity, and performance issues. We examined the utilized algorithm and the application interface to maximize the overall performance and make a user-friendly application. Therefore, our crowdsourced-based tracing system will help the government identify and aware people in the fastest way and considerably reduce the infection rate.

IV. CONCLUSION

The number of coronaviruses confirmed test cases around the world is increasing at an alarming rate. The leading cause of this problem is that the infected people do not maintain their quarantine properly and walk outside without any concern. Thus, the virus quickly spreads out to another human body and affects the nearby person. Therefore, to stop the spread of this virus, this paper proposed a crowdsourced-based contact tracing framework. By utilizing the Haversine and circle formula, the unsafe area has been calculated. The algorithm has been tested for several test cases and datasets to validate its efficiency in polynomial time. Once the government identified the marked area, they will notify people staying in that significant zone. User ratings also verify the application scalability, security, and effectiveness.

Whereas previously launched applications in Bangladesh required Bluetooth connectivity, this novel app can perform smoothly without Bluetooth and Wi-Fi. This application will help the government reduce the virus spreading into human bodies and help people lead a healthy life again. Moreover, this work can be extended by adding some new features for the coronavirus. In the future, we will merge an incentive and penalizing model to ensure more participation of the citizens with some new features.

ACKNOWLEDGMENT

The authors are grateful to anonymous reviewers and editors for the enormous efforts.

REFERENCES

[1] Y. Tong, Z. Zhou, Y. Zeng, L. Chen, and C. Shahabi, "Spatial crowdsourcing: a survey," *The VLDB Journal*, vol. 29, no. 1, pp. 217–250, 2020.
[2] L. Islam, S. T. Alvi, M. N. Uddin, and M. Rahman, "Obstacles of mobile crowdsourcing: A survey," in *2019 IEEE Pune Section International Conference (PuneCon)*, pp. 1–4, IEEE, 2019.
[3] S. O'dea, "Number of smartphone users worldwide from 2016 to 2021," *Statista Research Department*, 2020.
[4] S. Zhu, Z. Cai, H. Hu, Y. Li, and W. Li, "zkcrowd: a hybrid blockchain-based crowdsourcing platform," *IEEE Transactions on Industrial Informatics*, vol. 16, no. 6, pp. 4196–4205, 2019.
[5] H. D. Das, R. Ahmed, N. Smrity, and L. Islam, "Bdonor: A geo-localised blood donor management system using mobile crowdsourcing," in *2020 IEEE 9th International Conference on Communication Systems and Network Technologies (CSNT)*, pp. 313–317, IEEE, 2020.
[6] Z. Y. Zu, M. D. Jiang, P. P. Xu, W. Chen, Q. Q. Ni, G. M. Lu, and L. J. Zhang, "Coronavirus disease 2019 (covid-19): a perspective from China," *Radiology*, vol. 296, no. 2, pp. E15–E25, 2020.
[7] P. Bahl, C. Doolan, C. De Silva, A. A. Chughtai, L. Bourouiba, and C. R. MacIntyre, "Airborne or droplet precautions for health workers treatingcovid-19?," *The Journal of Infectious Diseases*, 2020.
[8] R. Humphries, M. Spillane, K. Mulchrone, S. Wieczorek, M. O'Riordain, and P. Hövel, "A metapopulation network model for the spreading of sars-cov-2: Case study for Ireland," *Infectious Disease Modelling*, vol. 6, pp. 420–437, 2021.
[9] M. P. Crayne, "The traumatic impact of job loss and job search in the aftermath of COVID-19.," *Psychological Trauma: Theory, Research, Practice, and Policy*, vol. 12, no. S1, p. S180, 2020.
[10] H. Stevens and M. B. Haines, "Tracetogether: Pandemic response, democracy, and technology," *East Asian Science, Technology and Society: An International Journal*, vol. 14, no. 3, pp. 523–532, 2020.
[11] M. Zens, A. Brammertz, J. Herpich, N. Südkamp, and M. Hinterseer, "App-based tracking of self-reported covid-19 symptoms: analysis of questionnaire data," *Journal of medical Internet research*, vol. 22, no. 9, p. e21956, 2020.
[12] R. Mallik, D. Sing, and R. Bandyopadhyay, "GPS tracking app for police to track ambulances carrying COVID-19 patients for ensuring safe distancing," *Transactions of the Indian National Academy of Engineering*, vol. 5, pp. 181–185, 2020.
[13] S. Anwar, M. Nasrullah, and M. J. Hosen, "Covid-19 and Bangladesh: Challenges and how to address them," *Frontiers in Public Health*, vol. 8, 2020.
[14] E. González-González, G. Trujillo-de Santiago, I. M. Lara-Mayorga, S. O. Martinez-Chapa, and M. M. Alvarez, "Portable and accurate diagnostics for covid-19: Combined use of the miniPCR thermocycler and a well-plate reader for sars-cov-2 virus detection," *PloS one*, vol. 15, no. 8, p. e0237418, 2020.
[15] M. N. K. Boulos and E. M. Geraghty, "Geographical tracking and mapping of coronavirus disease COVID-19/severe acute respiratory syndrome coronavirus 2 (SARS-COV-2) epidemic and associated events around the world: how 21st century GIS technologies are supporting the global fight against outbreaks and epidemics," *International Journal of Health Geographics*, vol. 19, no. 8, pp. 1–12, 2020.
[16] M. Rezaei, A. A. Nouri, G. S. Park, and D. H. Kim, "Application of geo-graphic information system in monitoring and detecting the COVID-19 out-break," *Iranian Journal of Public Health*, vol. 49, no. 1, pp. 114–116, 2020.
[17] J. Ammendolia, J. Saturno, A. L. Brooks, S. Jacobs, and J. R. Jambeck, "An emerging source of plastic pollution: environmental presence of plastic personal protective equipment (PPE) debris related to COVID-19 in a metropolitan city," *Environmental Pollution*, vol. 269, p. 116160, 2021.
[18] S. Hossain, M. N. Hasan, M. N. Islam, M. R. Mukto, M. S. Abid, and F. Khanam, "Information-based mobile application to tackle covid-19 circumstances," *Journal of Scientific Research and Reports*, vol. 27, no. 1, pp. 78–92, 2021.
[19] E. Shim, A. Tariq, W. Choi, Y. Lee, and G. Chowell. "Transmission potential and severity of covid-19 in south korea", *International Journal of Infectious Diseases*, vol. 93, pp. 339–344, 2020.
[20] M. N. Islam, S. R. Khan, N. N. Islam, M. Rezwan-A-Rownok, S. R. Zaman, and S. R. Zaman, "A mobile application for mental health care duringcovid-19 pandemic: Development and usability evaluation with system usability scale," in *International Conference on Computational Intelligence in Information System*, pp. 33–42, Springer, 2021.
[21] Y. Jung and R. Agulto, "A public platform for virtual IoT-based monitoring and tracking of covid-19," *Electronics*, vol. 10, no. 1, p. 12, 2021.
[22] T. Alanzi, "A review of mobile applications available in the app and google play stores used during the covid-19 outbreak," *Journal of Multidisciplinary Healthcare*, vol. 14, pp. 45-57, 2021.
[23] S. S. Hassan, S. D. Bibon, M. S. Hossain, and M. Atiquzzaman, "Security threats in Bluetooth technology," *Computers & Security*, vol. 74, pp. 308–322, 2018.
[24] Y. Meng, J. Li, H. Zhu, X. Liang, Y. Liu, and N. Ruan, "Revealing your mobile password via wifi signals: Attacks and countermeasures," *IEEE Transactions on Mobile Computing*, vol. 19, no. 2, pp. 432–449, 2019.
[25] S. Brack, L. Reichert, and B. Scheuermann, "CAUDHT: Decentralized contact tracing using a DHT and blind signatures," in *2020 IEEE 45th Conference on Local Computer Networks (LCN)*, pp. 337–340, IEEE, 2020.
[26] A. M. Lonzetta, P. Cope, J. Campbell, B. J. Mohd, and T. Hayajneh, "Security vulnerabilities in bluetooth technology as used in IoT," *Journal of Sensor and Actuator Networks*, vol. 7, no. 3, p. 28, 2018.
[27] I. Shammugam, G. N. Samy, P. Magalingam, N. Maarop, S. Perumal, and B. Shanmugam, "Information security threats encountered by malaysian public sector data centers," *Indonesian Journal of Electrical Engineering and Computer Science*, vol. 21, no. 3, pp. 1820–1829, 2021.
[28] T. Alam and M. S. Rahman, "To trace or not to trace: Saving lives fromcovid-19 at the cost of privacy breach in Bangladesh," *Qatar Medical Journal*, vol. 2020, no. 3, 2020.